\documentclass[12pt]{iopart}
\usepackage{iopams}
\usepackage{amssymb, latexsym,graphics,wasysym}

\begin{document}

\title[Search Method for Gravitational Waves from Pulsar Glitches]
{An Evidence Based Time-Frequency Search Method for Gravitational Waves from Pulsar Glitches}
\author{James Clark,~Ik Siong Heng,~Matthew Pitkin,~Graham Woan}

\begin{abstract}
We review and expand on a Bayesian model selection technique for the detection of gravitational waves
from neutron star ring-downs associated with pulsar glitches.  The algorithm works with power spectral densities
constructed from overlapping time segments of gravitational wave data.  Consequently, the original approach was at
risk of falsely identifying multiple signals where only one signal was present in the data.  We introduce an
extension to the algorithm which uses posterior information on the frequency content of detected signals to
cluster events together.  The requirement that we have just one detection per signal is now met with the additional
bonus that the belief in the presence of a signal is boosted by incorporating information from adjacent time
segments.
\end{abstract}


\maketitle

\section{Introduction}
Pulsar glitches are characterised by a step increase in the pulsar's rotation frequency.  The mechanisms responsible
for pulsar glitches are unclear but there are two main candidates to explain the underlying process.  For older pulsars 
it seems glitches are likely caused by a dramatic decoupling between the star's solid crust and 
superfluid interior~\cite{Cheng:1988} while glitches in younger pulsars may be associated with reconfigurations of the crust as spin-down reduces the centrifugal force and the crust reaches breaking strain~\cite{Franco:2000}.  In either case, 
the disruption should excite oscillatory modes throughout the neutron star.

The excitation of quadrupolar quasinormal modes in a neutron star leads to the emission of a short,
distinctive burst of gravitational radiation in the form of a decaying sinusoid, or `ring-down'~\cite{Thorne:1969},
with typical frequencies and decay times in the range of $1-3~{\rm kHz}$ and $50-500~{\rm ms}$, respectively.

In \cite{Clark:2007} we present a search method based on Bayesian model selection where the evidence 
for a ring-down signal is compared with the evidence for a noise model, taken to be Gaussian white noise for simplicity.
In this work we review this approach and demonstrate how it may be modified to cluster odds ratios
computed from power spectral densities in the time-signal frequency plane.  In essence, we use the frequency posterior
from the most probable event in a cluster to set a prior to compute the odds ratios in adjacent time-frequency pixels.

\section{Bayesian Model Selection}
Given some data $D$, a model $M$ and some background information $I$, representing
some prior knowledge of the system, we can write down the posterior probability
for the model using Bayes' theorem,

\begin{equation}\label{eq:bayes}
p(M|D,I)=\frac{p(M|I)p(D|M,I)}{p(D|I)},
\end{equation}

where $p(M|I)$ is the prior probability of the model, $p(D|M,I)$ is the likelihood
or \emph{evidence} of the model $M$ and $p(D|I)$ is the probability of the data.  We are 
free to assign any suitable prior for the model and the model likelihood or \emph{evidence} 
is computed by marginalising the likelihood of the data over all of the model
parameters, ${\underline \theta}$,

\begin{equation}\label{eq:evidence}
p(D|M,I)=\int_{\underline \theta} p({\underline \theta}|M,I)p(D|{\underline \theta},M,I)~d{\underline \theta}.
\end{equation}

We wish to address the question, `does the data $D$ contain a ring-down gravitational wave or
is the data simply Gaussian white noise?'.  Let $M_s$ denote the model that the data contains a ring-down
signal $S$ and $M_n$ denote the model that the data contains only noise.  We then write down the `odds ratio'
$\Omega_{s,n}$ which is the ratio of the posterior probabilities for each model,

\begin{equation}
\Omega_{s,n} = \frac{p(M_s|D,I)}{p(M_n|D,I)}.
\end{equation}

Substituting Bayes' theorem (equation \ref{eq:bayes}) for the posteriors, we see that the odds ratio may be
divided into a term independent of the data, (the \emph{prior odds}) and a term dependent on
the data (the \emph{Bayes factor}):

\begin{equation}
\Omega_{s,n} = \frac{p(M_s|I)}{p(M_n|I)}\times\frac{p(D|M_s,I)}{p(D|M_n,I)}.
\end{equation}

In practice, the prior odds can be assigned by computing the Bayes factor (i.e., evidence ratio) in some sample
of data away from the pulsar glitch trigger.  If the prior odds are then set equal to the reciprocal of this 
off-source Bayes factor, the odds ratio in the on-source data will rise above unity when there is greater
evidence for a ring-down than there was in the off-source data.  However, it is necessary to set a higher threshold
than unity to claim a detection since $\Omega_{s,n}=1$ simply indicates no preference for either model.
The threshold required for a given level of confidence is determined empirically through
simulated signal injections.  This is demonstrated and explored in more detail in \cite{Clark:2007} and we do not 
consider the issue here.

\begin{figure}
\centering
\begin{tabular}{ll}

\begin{minipage}{0.5\linewidth}
\includegraphics
{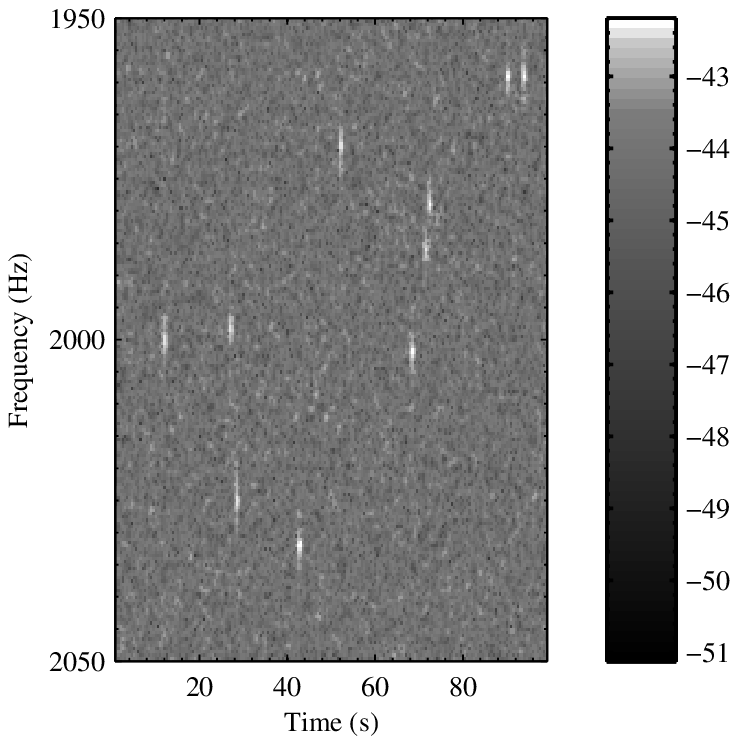}
\end{minipage}

&

\begin{minipage}{0.5\linewidth}
\includegraphics
{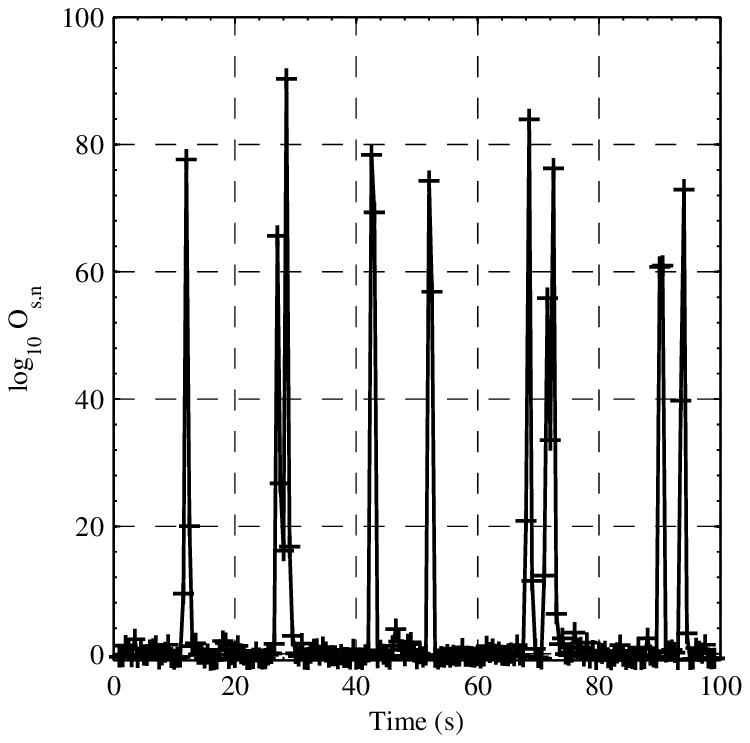}
\end{minipage}

\\

\end{tabular}
\caption{
\emph{Left panel}: Spectrogram with 10 randomly placed ring-down signals in synthetic Gaussian white noise
of amplitude spectral density $10^{-1/2}~{\rm Hz}$.  The signals all share an initial amplitude $h_0=5\times
10^{-21}$ and decay time $\tau=0.2~{\rm s}$.
\emph{Right panel}: The base ten logarithm of the odds in favour of a ring-down signal versus Gaussian white noise
in each time bin of the spectrogram in the left panel.  Notice that each signal spills into neighbouring time bins,
generating multiple excess odds.
}
\label{fig:injectogram}
\end{figure}

To demonstrate the detection process we inject 10 ring-down waveforms with randomly chosen frequencies in the range
$1950~{\rm Hz}$ to $2050~{\rm Hz}$, decay time
$\tau=0.2~{\rm s}$ and initial amplitude $h_0=5\times10^{-21}$ into $100~{\rm s}$ of synthetic Gaussian white noise with
amplitude spectral density $\sim 10^{-22}~{\rm Hz}$.
The spectrogram in the left panel of figure~\ref{fig:injectogram} shows the injected signals in the 
time-frequency plane.  The odds ratio $\Omega_{s,n}$ in each time bin of the spectrogram is shown in the right panel of
figure \ref{fig:injectogram}.  The prior odds in this example are computed from sythetic Gaussian white noise with no
injections.

As expected, the injected signals cause excesses in the odds ratio, indicating a preference for the
signal model over the noise model.  Due to the overlap ($75\%$) between the time segments used to construct the spectrogram,
each signal actually appears in multiple time bins, causing multiple excesses in the odds ratio.  In the following
section we describe a basic algorithm to handle these multiple excesses.

\section{Odds Clustering Algorithm}
The multiple integral for the evidence computation (equation \ref{eq:evidence}) is computationally expensive to evaluate.
In practice, it quickly becomes necessary to divide the parameter space into manageable segments which can be distributed
out as individual tasks on large scale computing clusters.  The results can then simply be summed together to reconstruct
the evidence computed over the entire prior range.  Equation \ref{eq:splitevidence} describes the division of the parameter
space,

\begin{eqnarray}\label{eq:splitevidence}
p(D|M_s) & = & \int_{h_0} \int_{\tau} \int_{\omega_0} ~p(h_0,\tau,\omega_0|M_s)p(D|h_0,\tau,\omega_0,M_s)~dh_0 d\tau d\omega_0 \nonumber \\
	& = & \int_{h_0} \int_{\tau} \int_{\omega_0(a)}^{\omega_0(b)} ~p(h_0,\tau,\omega_0|M_s)p(D|h_0,\tau,\omega_0,M_s)~dh_0 d\tau d\omega_0 \nonumber \\
	& + & \int_{h_0} \int_{\tau} \int_{\omega_0(b)}^{\omega_0(c)} ~p(h_0,\tau,\omega_0|M_s)p(D|h_0,\tau,\omega_0,M_s)~dh_0 d\tau d\omega_0 \nonumber \\
	& + & ...
\end{eqnarray}

where $p(h_0,\tau,\omega_0|M_s)$ is the prior on the ring-down parameters and $p(D|h_0,\tau,\omega_0,M_s)$ is the likelihood
of the data given the parameters.  The integral limits $[\omega_0(a),\omega_0(b)]$ and $[\omega_0(b),\omega_0(c)]$ correspond
the the distribution of the parameter space over a computing cluster.

\begin{figure}
\centering
\begin{tabular}{ll}

\begin{minipage}{0.5\linewidth}
\includegraphics
{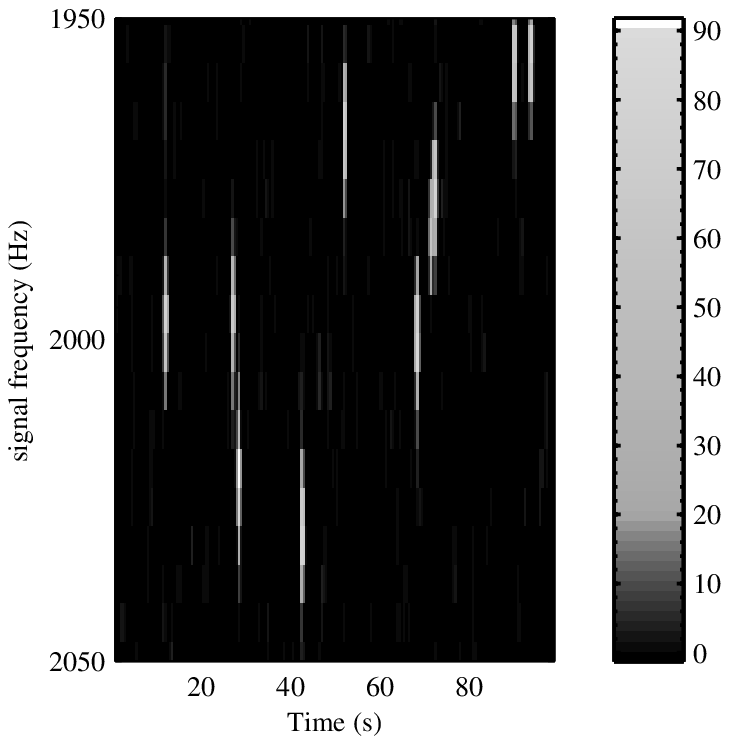}
\end{minipage}

&

\begin{minipage}{0.5\linewidth}
\includegraphics
{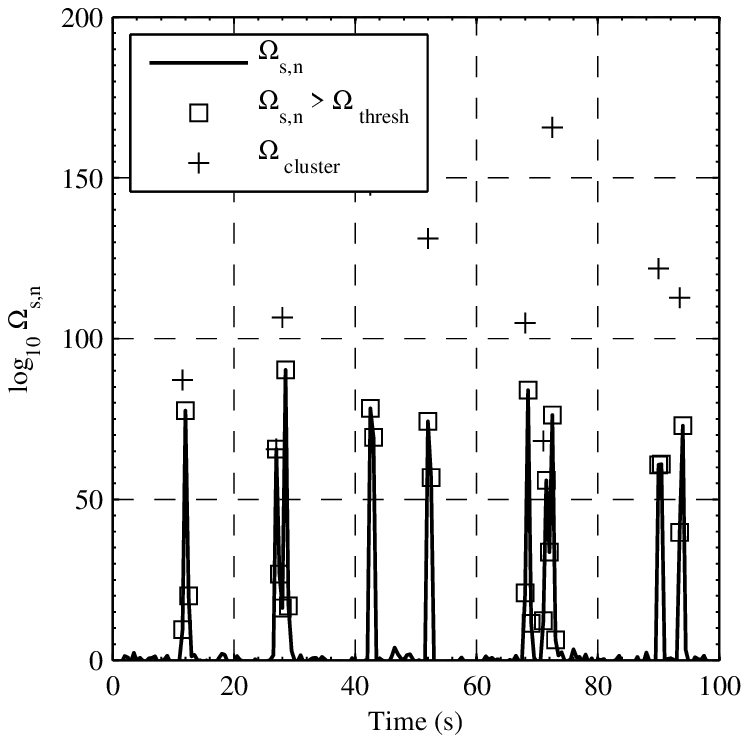}
\end{minipage}

\\

\end{tabular}
\caption{
\emph{Left panel}:  The log odds (`lodds') in the time-signal frequency plane   We term this representation
a `loddogram'.
\emph{Right panel}:  The log odds in each spectrogram time bin.  Time bins in which the odds are greater
than the detection threshold $\Omega_{\rm thresh}$ are marked with a square.  The output of the clustering algorithm
is marked by crosses.  We see that without the clustering algorithm, there would be more signal `detections' than
there are signals.
}
\label{fig:speclods}
\end{figure}

The first line in the equation above shows the evidence integral from equation \ref{eq:evidence} with the frequency space divided
into seperate segments on the lines which follow.  Each of the integrals for the divided parameter space then represents the individual computational tasks, or `jobs'.  More interestingly, this division allows us to construct a time-frequency map of
odds ratios.  Since we typically deal with the base 10 logarithm of the odds ratio, we adopt the term \emph{loddogram} to describe
the time-frequency odds map. The frequency resolution here is given by the width of the frequency prior for each job (e.g., $\omega_0(b)-\omega_0(a)$)
.  The left panel of figure~\ref{fig:speclods} shows the loddogram produced by the synthetic data.  The frequency resolution for each job in this example is $5~{\rm Hz}$.  Unsurprisingly, the loddogram closely resembles the spectrogram from figure~\ref{fig:injectogram} and all ten injections are clearly visible.

To cluster events together and maximise the information available, we begin by identifying odds values above a threshold
$\Omega_{\rm thresh}$.  This picks out time bins where the bulk of the signal power lies and
yields the time stamp assigned to the event.  The algorithm then goes through each threshold crossing and locates the loddogram pixels with
the highest odds values.  These are used to construct new (flat) priors for the frequency of the signal which caused each threshold
crossing.  The new priors are then used to compute the odds $o_{s,n}$ in favour of a ring-down signal in the time bins adjacent to the initial threshold
crossings.  A second threshold $o_{\rm thresh}$ is then applied.  If $o_{s,n} > o_{\rm thresh}$ then we add the logarithm of
the odds calculated from the narrow prior (i.e, $\log_{10} o_{s,n}$) to the logarithm of the odds value which triggered the cluster formation
in the first place ($\log_{10} \Omega_{s,n}$).  This way, we are effectively taking the joint probability of multiple hypotheses in adjacent
time bins.  One outstanding issue here is, of course, the fact that these probabilities are strongly correlated due to the overlap between
time bins.  For now, we will neglect this correlation in favour of demonstrating the general approach. More succintly, the final value
of the odds ratio for each odds cluster at time $T$ is,

\begin{eqnarray}
& \log_{10} \Omega_{\rm cluster}(T) = \log_{10} \Omega_{\rm s,n}(T) + \sum_i^n \log_{10} o_{\rm i} \nonumber \\ 
& \Leftrightarrow \Omega_{s,n} > \Omega_{\rm thresh}~\&~o_{i}>o_{\rm thresh},
\end{eqnarray}

where $\Omega_{s,n}$ is the odds ratio constructed from the full prior range, $o_{i}$ is the odds ratio
constructed from the prior given by the strongest loddogram pixel in $\Omega_{s,n}(T)$ and $n$ is the number
of consecutive time bins in which $o_i>o_{\rm thresh}$.

The right panel of figure~\ref{fig:speclods} shows the results of applying this clustering algorithm.  The solid line shows the original
time series of odds ratios, squares show odds ratios above the initial detection threshold $\Omega_{\rm thresh}$ and, finally, the odds ratio
in each cluster $\Omega_{\rm cluster}$ is indicated by a cross.  We see that the clustering algorithm correctly distinguishes between the injections
and, for those signals which satisfy the condition $o_i>o_{\rm thresh}$, the clustered odds value is significantly higher than the initial
threshold crossing (the square symbols).

\section{Conclusion}
In these proceedings we have given a brief overview of the model selection technique for gravitational wave
detection introduced in \cite{Clark:2007}. We have expanded on the original work and introduced a simple clustering algorithm which only associates
a single odds value with a single signal.  This is an improvement over the original work where overlap between Fourier
transform time segments could potentially result in multiple `detections' for a single signal.  Further, the
posterior frequency information from time bins where the odds ratio crosses some detection threshold is used to construct
a new prior for computing the odds ratio in adjacent time bins.  If this secondary odds ratio crosses its respective
threshold, we compute the joint probability in favour of a signal across several time bins.

It should be noted that we do not consider any kind of sensitivity or performance estimate in this work as we wish
only to highlight a potential approach.  Furthermore, the computation of a joint probability between overlapping time
bins should formally account for correlations.  We will consider these factors in future work and in the application
to real gravitational wave data.

\end{document}